\documentclass{nature}

\usepackage{graphicx}

\makeatletter
\let\saved@includegraphics\includegraphics
\AtBeginDocument{\let\includegraphics\saved@includegraphics}
\renewenvironment*{figure}{\@float{figure}}{\end@float}
\makeatother

\usepackage[colorlinks=true,linkcolor=blue,citecolor=blue,urlcolor=blue]{hyperref}
\usepackage{authblk}
\usepackage{amsmath}
\usepackage{float}
\usepackage{siunitx}
\usepackage{xr}
\usepackage{cleveref}
\usepackage[dvipsnames]{xcolor}
\usepackage{ulem}
\DeclareSIUnit\rydberg{Ry}
\DeclareSIUnit\atomicunit{a.u.}
\DeclareSIUnit\bohr{\text{\ensuremath{a_0}}}

\newcommand{\VEC}[1]{\mathbf{#1}}

\begin{document}

\title{Electrical engineering of topological magnetism in two-dimensional heterobilayers}

\author[1,2,3, *]{Nihad Abuawwad}

\author[4]{ Manuel dos Santos Dias}

\author[3]{Hazem Abusara}

\author[1,2,*]{Samir Lounis}

\affil[1]{Peter Gr\"unberg Institut and Institute for Advanced Simulation, Forschungszentrum J\"ulich \& JARA, 52425 J\"ulich, Germany}
\affil[2]{Faculty of Physics, University of Duisburg-Essen and CENIDE, 47053 Duisburg, Germany}
\affil[3]{Department of Physics, Birzeit University, PO Box 14, Birzeit, Palestine}
\affil[4]{ Scientific Computing Department, STFC Daresbury Laboratory, Warrington WA4 4AD, United Kingdom}
\affil[*]{n.abuawwad@fz-juelich.de; s.lounis@fz-juelich.de}

\maketitle

\begin{abstract}

The emergence of topological magnetism in two-dimensional (2D) van der Waals (vdW) magnetic materials promoted 2D heterostructures as key building-blocks of devices for information technology based on topological concepts. 
Here, we demonstrate the all-electric switching of the  topological nature of individual magnetic objects emerging in 2D vdW heterobillayers.
We show from the first principles that an external electric field modifies the vdW gap between CrTe $_2$ and (Rh, Ti)Te$_2$ layers and alters the underlying magnetic interactions.
This enables switching between ferromagnetic skyrmions and meron pairs in the CrTe$_2$/RhTe$_2$ heterobilayer while it enhances the stability of frustrated antiferromagnetic merons in the CrTe$_2$/TiTe$_2$ heterobilayer. 
We envision that the electrical engineering of distinct topological magnetic solitons in a single device could pave the way for novel energy-efficient mechanisms to store and transmit information with applications in spintronics.

\end{abstract}

\maketitle

\section*{Introduction}

Two-dimensional van der Waals materials display a range of remarkable physical properties that can be altered either by incorporating them into heterostructures or through the application of external forces. The recent uncovering of extended ferromagnetic ordering ~\cite{cri3, cr2ge2te2} at the atomic layer level introduces a new avenue for designing 2D materials and their combined structures, enhancing their potential for spintronics, valleytronics, and magnetic tunnel junction switches ~\cite{Gibertini2019, Gong2019, McGuire2020, Jiang2021b, rev15, rev16, rev17, rev18}.
Often, these 2D materials display straightforward magnetic orders like ferromagnetic (FM) or antiferromagnetic (AFM). 
Yet, they can also showcase intricate noncollinear magnetic patterns and support topological magnetic conditions, such as ferromagnetic skyrmions ~\cite{noncol-1, noncol-2, noncol-3}.
FM skyrmions are topologically protected magnetic textures characterized by a unique swirling configuration of magnetic moments or spins. 
Their small size, stability against perturbations, and the ability to be moved with minimal current make them promising candidates for future data storage and spintronic devices. 
These chiral magnetic states typically arise due to the interplay between the Heisenberg exchange and the relativistic Dzyaloshinskii-Moriya interaction (DMI)~\cite{Moriya, dzy} in noncentrosymmetric materials with significant spin-orbit coupling.
In 2D materials, skyrmions form when the overall magnetization is perpendicular to the plane of the magnetic layer.
Conversely, merons are created when the magnetization lies within the plane. 
Skyrmions possess an integer topological charge, whereas merons have a half-integer charge.
Thus, these are two types of chiral spin-textures with marked differences in their properties~\cite{Goebel2021, afm-meron, afm-meron-1}.

In the context of 2D heterostructures, Néel-type FM skyrmions have been experimentally observed in Fe$_3$GeTe$_2$~\cite{sky-0, sky-1, sky-2, sky-3, sky-4}, and their theoretical existence has been proposed in many systems like MnXTe (X = S and Se)~\cite{Liang2020}, CrInX3 (X = Te, Se)~\cite{Yuan2020}, bilayer Bi$_2$Se$_3$-EuS~\cite{Nogueira2018}, and CrTe$_2$/(Ni, Zr, Rh)Te$_2$ heterobilayers~\cite{nihad}. 
However, FM merons remain undetected in 2D van der Waals (vdW) heterostructures, with their identification limited to more conventional thin films and disks~\cite{m1, m2, m4, m5}.
Notably, merons have been theoretically predicted in free-standing monolayers of CrCl$_{3}$~\cite{Lu2020}.
Additionally, our recent theoretical work suggests the potential presence of Néel-type frustrated AFM merons in both isolated CrTe$_2$ monolayers and CrTe$_2$/(Ti, Nb, Ta)Te$_2$ heterobilayers~\cite{nihad}.
This topological state can be understood
by considering each AFM sublattice separately, and results from diverse pairwise combinations of ferromagnetic meronic textures (meron-meron, meron-antimeron, or antimeron-antimeron)~\cite{amal}.
Consequently, various values for the overall topological charge $Q$ are possible, determined by the topological charges (q) in the three distinct AFM sublattices.

While the possibility of manipulating such topological states through electric fields is recognized, the few experimental studies exploring electric-field-induced switching of skyrmions and skyrmion bubbles considered transition-metal multilayers~\cite{hsu2017electric, ma2018electric, D0NA00009D} and multiferroic heterostructures~\cite{wang2020electric}.
On the theoretical side, research has focused on the alterations to either the magnetic anisotropy (MA) or the DMI~\cite{fook2016gateable, nakatani2016electric}, whether influenced directly by the electric field or indirectly due to the strain that it effects.
More recently, simulations indicated the possibility of stabilizing individual magnetic skyrmions in an ultrathin transition-metal film  via external electric fields through the combined effect of the exchange interaction, the DMI, and the MA that dictates the characteristics of magnetic skyrmions~\cite{electric}.
A nonequilibrium approach was also demonstrated, with the DMI being generated by a femtosecond electric field pulse in an ultrathin metal film ~\cite{PhysRevB.104.L060409}.
However, no predictions have so far been made on whether skyrmions can be transformed into other topological magnetic states. 

\begin{figure}[H]
\centering
   \includegraphics[width=\textwidth]{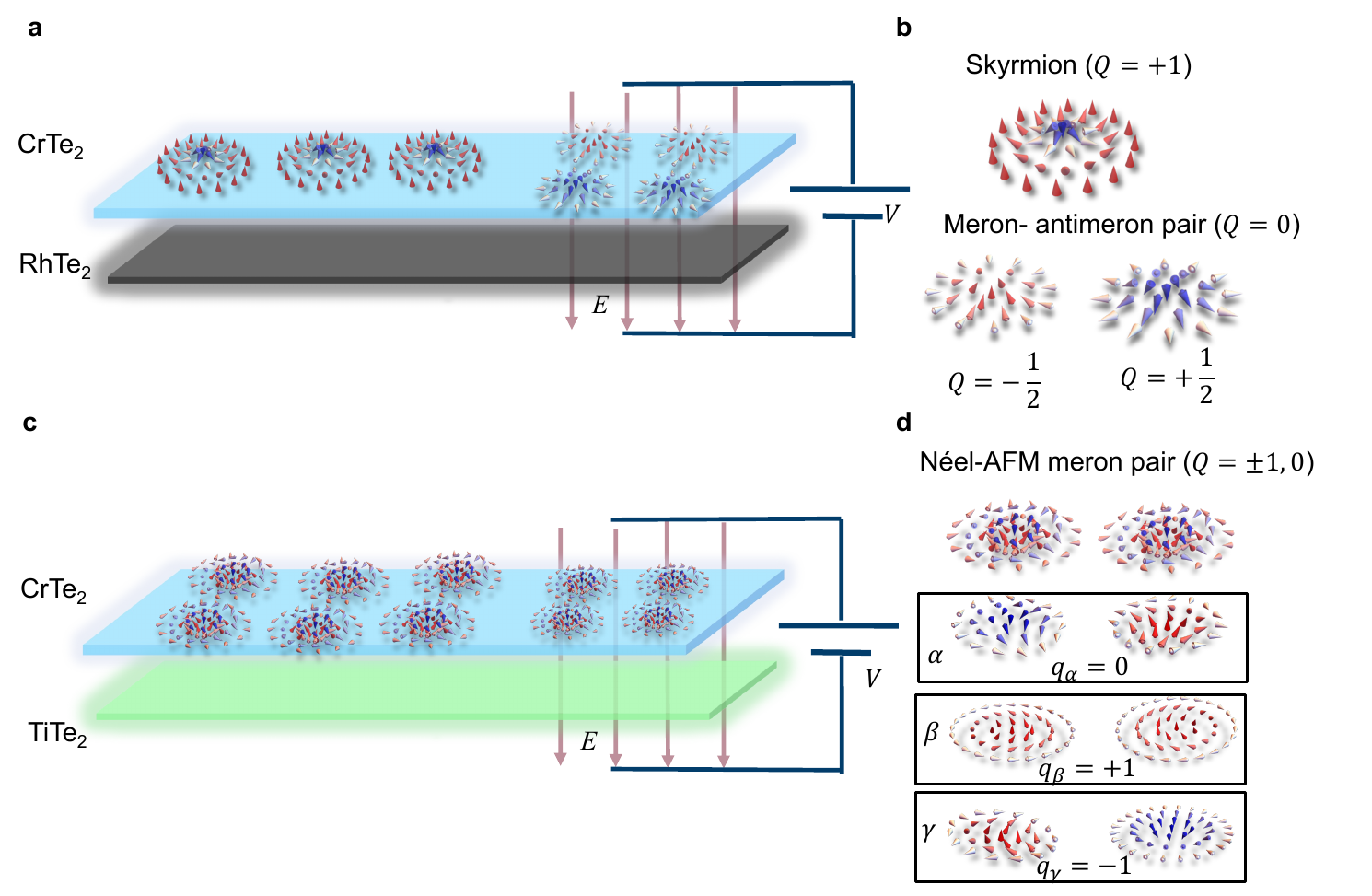}
    \caption{\label{figure_0}
    \textbf{Overview of the magnetic topological states with and without an electric field in heterobilayers.}
    \textbf{a} CrTe$_2$/RhTe$_2$ heterobilayer, with
    \textbf{b} showing FM skyrmions and merons and their net topological charges ($Q$) which arise in this bilayer.
    \textbf{c} CrTe$_2$/TiTe$_2$ heterobilayer, with
    \textbf{d} showing N\'eel AFM merons which are composed from three FM merons and/or antimerons living in different sublattices ($\alpha$, $\beta$, and $\gamma$), with distinct sublattice topological charges ($q$).
    For the shown example, the N\'eel AFM meron pair has $Q=0$, while the sublattice topological charges are all different ($q_\alpha = 0$, $q_\beta = 1$, $q_\gamma = -1$).
    }
\end{figure}

In this study, we use atomistic spin models in combination with first-principles calculations to uncover the non-trivial impact of the electric field on noncollinear magnetic structures in heterobilayers with CrTe$_2$ as the magnetic layer.
For the CrTe$_2$/RhTe$_2$ bilayer sketched in Fig.~\ref{figure_0}a, we discover all-electrical switching between two topologically different magnetic structures, FM skyrmions and FM meron pairs (Fig.~\ref{figure_0}b).
The perpendicular electric field has a strong influence on the interlayer spacing between the 2D materials, which modifies several key magnetic interactions: the Heisenberg exchange interaction, the DMI and the MA.
These electric-field-induced alterations enable the transition of skyrmions into meron structures and vice-versa.
A very different scenario arises when interfacing CrTe$_2$ with TiTe$_2$, sketched in Fig.~\ref{figure_0}c, which leads to the emergence of frustrated AFM merons (Fig.~\ref{figure_0}d) whose stability and size can be tuned by the applied electric field. 
Our findings provide a foundation for further exploration in electrically tunable magnetic systems, offering innovative avenues for the design and control of novel spintronic functionalities.

\section*{Results}

\subsection{Electric field modulation of the electronic properties in heterobilayers.}

\begin{figure}[t!]
\centering
   \includegraphics[width=\textwidth]{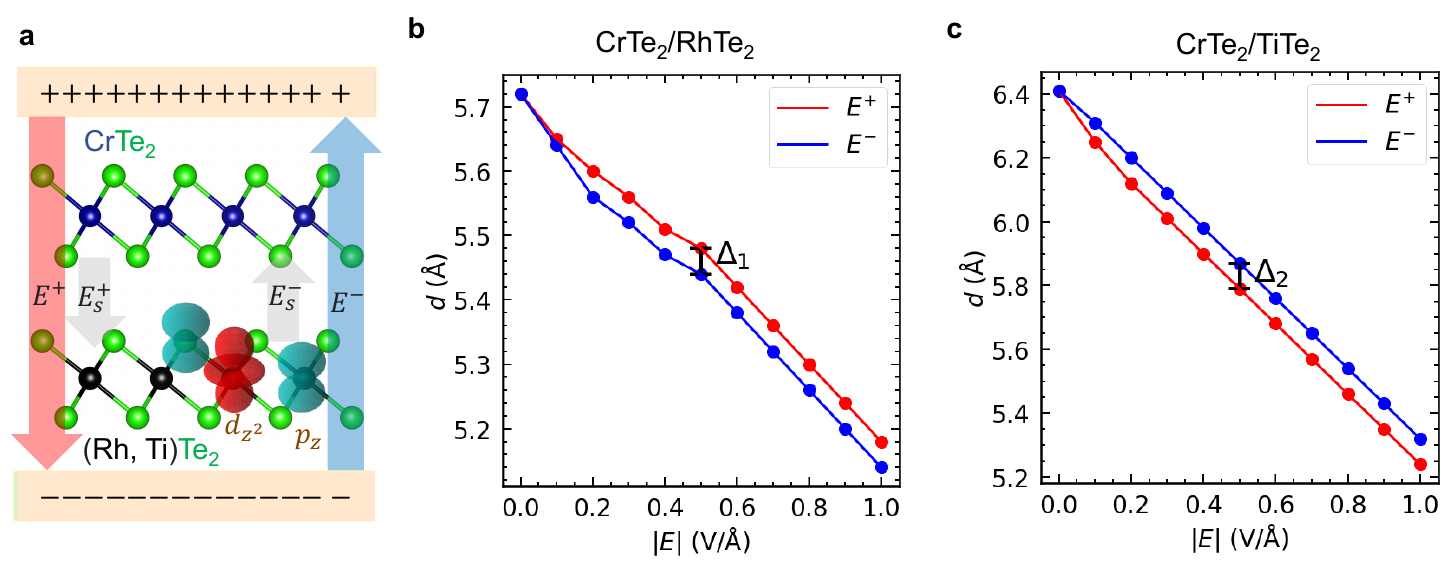}
    \caption{\textbf{Impact of the perpendicular electric field on the interlayer distances.}
    \textbf{a} CrTe$_2$/(Rh, Ti)Te$_2$ heterobilayers in a capacitive environment.
    $E_\mathrm{s}^{+}$ ($E_\mathrm{s}^{-}$) are the screened electric fields due to the positive (negative) applied fields.
    The interlayer distances $d$ as a function of the magnitude of the applied electric field for \textbf{b} CrTe$_2$/RhTe$_2$ and \textbf{c} CrTe$_2$/TiTe$_2$ heterobilayers, respectively.
    $\Delta_{1}$ and $\Delta_{2}$ indicate the difference in the interlayer distance for applied fields of equal magnitude in opposite directions for the respective bilayers.}
    \label{figure_1} 
\end{figure}

We first performed calculations utilizing Quantum Espresso to obtain the structural and electronic properties and how they respond to an applied electric field.
Fig.~\ref{figure_1}a illustrates the setup used to investigate the effects of perpendicular electric fields of both polarities on the interlayer distances of CrTe$_2$/(Rh, Ti)Te$_2$ heterobilayers.
In both cases, we found that the effect of the applied electric field is to decrease the interlayer distance, as shown in as shown in Fig.~\ref{figure_1}b-c.
For instance, when applying a field of $+\SI{1.0} {V/\angstrom}$, the interlayer distance decreases from $\SI{5.72}{\angstrom}$ to $\SI{5.18}{\angstrom}$ in the CrTe$_2$/RhTe$_2$ heterobilayer, and from $\SI{6.41}{\angstrom}$ to $\SI{5.32}{\angstrom}$ in the CrTe$_2$/TiTe$_2$ heterobilayer.
The most striking aspect is the asymmetrical behavior of the interlayer distance when subjected to positive versus negative electric fields.
In the given example of $E = +\SI{1.0} {V/\angstrom}$, this asymmetry manifests as a difference between the respective interlayer distances of $\Delta_1=\SI{0.04}{\angstrom}$ in the CrTe$_2$/RhTe$_2$ heterobilayer and $\Delta_2=\SI{0.07}{\angstrom}$ in the CrTe$_2$/TiTe$_2$ heterobilayer. 
We note that such an asymmetry with respect to the polarity of the applied electric field was also found for the dielectric properties of a graphene/MoS2 van der Waals heterostructure~\cite{asy-mos2}.
This suggests that the electric field polarity plays a significant role in determining the behavior of these heterobilayers and warrants further investigation.

The broad effect of the applied electric field on the heterobilayer can be explained as follows.
The electric field polarizes each layer separately, and as the induced electric dipoles on each layer are parallel to each other this leads to an attraction between the two layers and to a reduction of the van der Waals gap between them.
This picture is validated by analysing and decomposing the charge density obtained from our DFT calculations, which results shown in Supplementary Fig.~\ref{supp-fig-0} for field values of $\SI{\pm0.5}{V/\angstrom}$ and $\SI{\pm1.0}{V/\angstrom}$.
Firstly, we found that the electric field does not transfer charge between the layers.
Secondly, the charge transfer within a layer follows the direction of the applied electric field, as expected.
The charge accumulating on one of the Te atoms comes both from the other Te atom and the transition metal atom, which shows that the induced electric polarization is not symmetric around the center of the layer (defined by the transition metal atom).
We also determined that the charge is transferred to/from the p$_z$ orbitals of the Te atoms and the d$_{z^2}$ orbital of the transition metals.
Thirdly, we also found that the applied electric field is strongly screened within the van der Waals gap, which we indicate by $E_s$ in Fig.~\ref{figure_1}.
The associated relative permittivity ($\varepsilon_r = E / E_s$) is in the 6 to 8 range for CrTe$_2$/RhTe$_2$ and in the 4 to 5 range for CrTe$_2$/TiTe$_2$, respectively.
Lastly, the magnitudes of the induced polarization of each layer and of the screened electric field are found to be asymmetric with respect to a change in polarity of the applied electric field.
This arises naturally from the heterogeneous nature of the bilayer, with each bilayer having a different electric polarizability, so that the voltage does not drop in the same way across each layer and hence the order in which the layers are arranged with respect to the polarity of the applied electric field matters.
The ensuing differences in the induced dipoles on each layer lead to a different attraction between the layers when the polarity is reversed, which explains the previously-found variation in the interlayer distance.
The change in the interlayer distance driven by the applied electric field shows a strong electroelastic coupling, which also has strong consequences for the magnetic properties.

\subsection{Electric field control of magnetism in CrTe$_2$/XTe$_2$ heterobilayers.}

\begin{figure}[H]
    \centering
    \includegraphics[width=\textwidth]{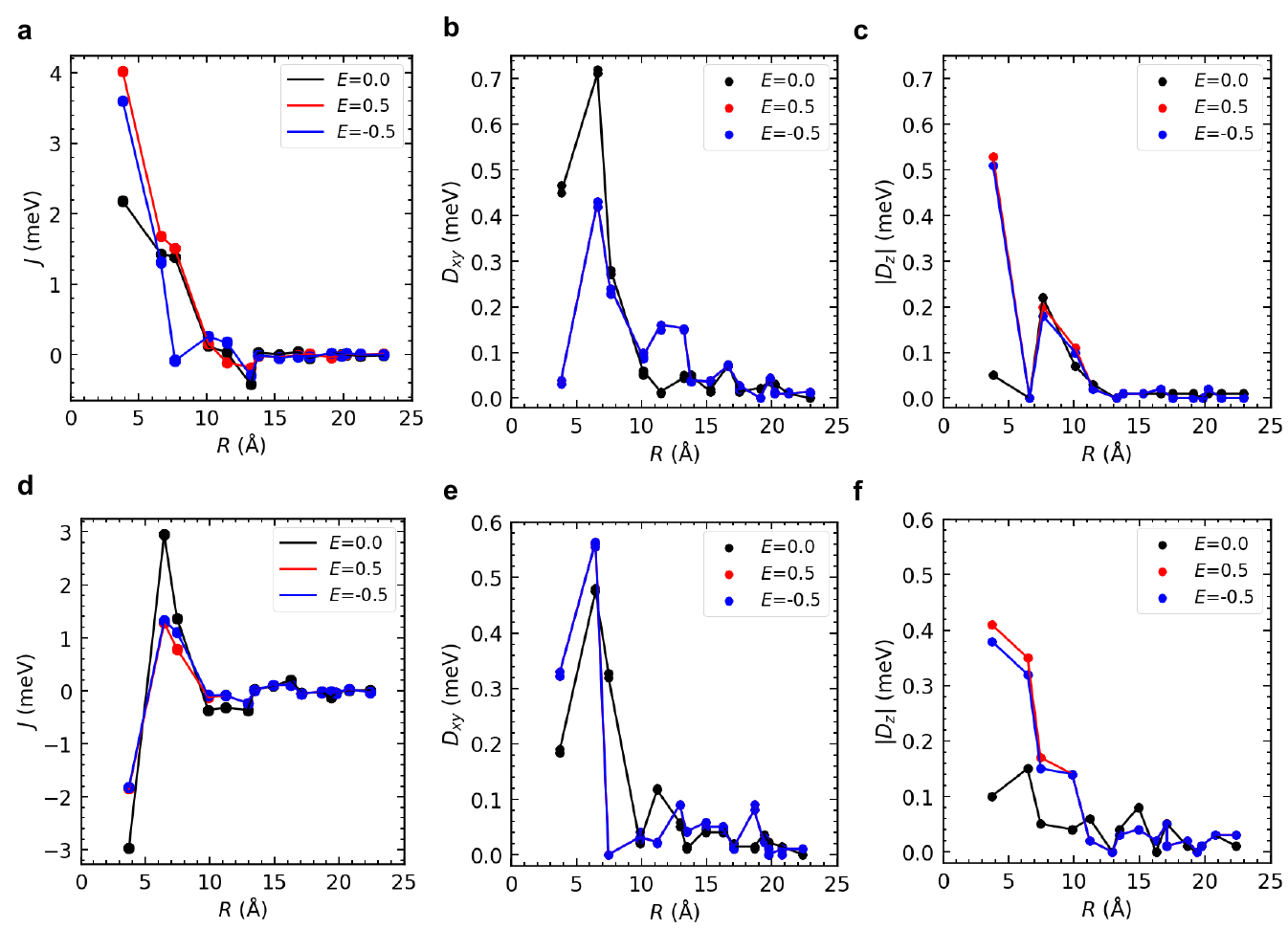}
    \caption{\label{figure_3}
    \textbf{Magnetic interactions as a function of the distance for the heterobilayers.}
    \textbf{a-c} CrTe$_2$/RhTe$_2$.
    \textbf{d-f} CrTe$_2$/TiTe$_2$.
    The symbols denote the Heisenberg exchange ($J$), and the in-plane ($D_{xy} = \sqrt{D_x^2 + D_y^2}$) and out-of-plane ($|D_z|$) components of the Dzyaloshinskii-Moriya interaction.
    The distance between Cr atoms for a given pair is labelled $R$.}
\end{figure}

To investigate the impact of the electric field on the magnetic properties, we performed calculations using JuKKR for the optimized structures obtained from Quantum Espresso.
We computed the magnetic anisotropy energy and extracted the tensor of magnetic interactions as a function of interatomic distances utilizing the infinitesimal rotation method~\cite{inf-rot} (see \nameref{sec:methods} section). 
We found strong changes in the magnetic interactions of the heterobilayers, evidencing a strong magnetoelectric coupling. 

We first recap the magnetic interactions without an applied electric field (data shown with black dots in Fig.~\ref{figure_3}).
The Heisenberg exchange interactions $J$ for the CrTe$_2$/RhTe$_2$ heterobilayer are predominantly FM, while they have mixed AFM/FM character for CrTe$_2$/TiTe$_2$.
The energetics of spiralling magnetic states can be readily obtained from their lattice Fourier transform, and are shown in Supplementary Fig.~\ref{supp-fig-3}.
For CrTe$_2$/RhTe$_2$ the energy minimum is found at $\Gamma$, indicating the expected FM ground state, while for CrTe$_2$/TiTe$_2$ the energy minimum is shifted away from the K point, signifying a spiralling ground state that locally resembles the triangular N\'eel AFM state.
The non-magnetic layer breaks the inversion symmetry of CrTe$_2$ and enables the DMI, which is found to be stronger in proximity to RhTe$_2$ than to TiTe$_2$.
The final ingredient is the in-plane MAE, $K = 1.70$ meV for CrTe$_2$/RhTe$_2$ and $K = 0.95$ meV for CrTe$_2$/TiTe$_2$.
Performing atomistic spin dynamics using all the obtained magnetic interactions in zero magnetic field with the Spirit code, we find the appearance of N\'eel-type skyrmionic domains for CrTe$_2$/RhTe$_2$ and AFM meron pairs in a spiralling N\'eel AFM background for CrTe$_2$/TiTe$_2$.
These results were presented in our prior work~\cite{nihad}.
We now turn to the effect of the applied electric field on the magnetic interactions, discussing results for $|E| = 0.5$ V/\AA\ for definiteness.

For CrTe$_2$/RhTe$_2$, we see from Fig.~\ref{figure_3}a that the electric field doubles the magnitude of the first-neighbor $J$ while it strongly suppresses the third-neighbor $J$ for negative polarity while having little impact for positive polarity.
The electric-field-induced modifications to the DMI are more long-ranged, as seen in Fig.~\ref{figure_3}b-c.
The first-neighbor DMI vector is rotated from in-plane to out-of-plane, the magnitude of the second-neighbor DMI is suppressed by about 40\% without rotation, and the magnitude of some further-neighbor DMIs is enhanced by the electric field.
Lastly, the MAE is decreased to 1.08 meV for positive polarity and increased to 1.82 meV for negative polarity, which correspond to changes of $-26\%$ and by $+7\%$ with respect to zero applied field, respectively.
Due to the combination of weakened in-plane components of the DMI together with strengthened FM Heisenberg exchange, the ground state becomes a more conventional in-plane FM which supports meron pairs.
For CrTe$_2$/TiTe$_2$, Fig.~\ref{figure_3}d shows that the electric field strongly weakens both first- (AFM) and second-neighbor (FM) Heisenberg exchange and suppresses longer-ranged AFM interactions, while it enhances and tilts out of the Cr plane the first- and second-neighbor DMI and suppresses the third-neighbor DMI, Fig.~\ref{figure_3}e-f.
Similar to what was found for the RhTe$_2$ case, the MAE is reduced to 0.78 meV for positive polarity and enhanced to 1.22 meV for negative polarity, which are variations by $-18\%$ and by $+28\%$ of the zero field values, respectively.
In both cases, we find that the MAE variation the most pronounced asymmetry with respect to changing the polarity of the electric field.

\begin{figure}[H]
    \centering
    \includegraphics[width=\textwidth]{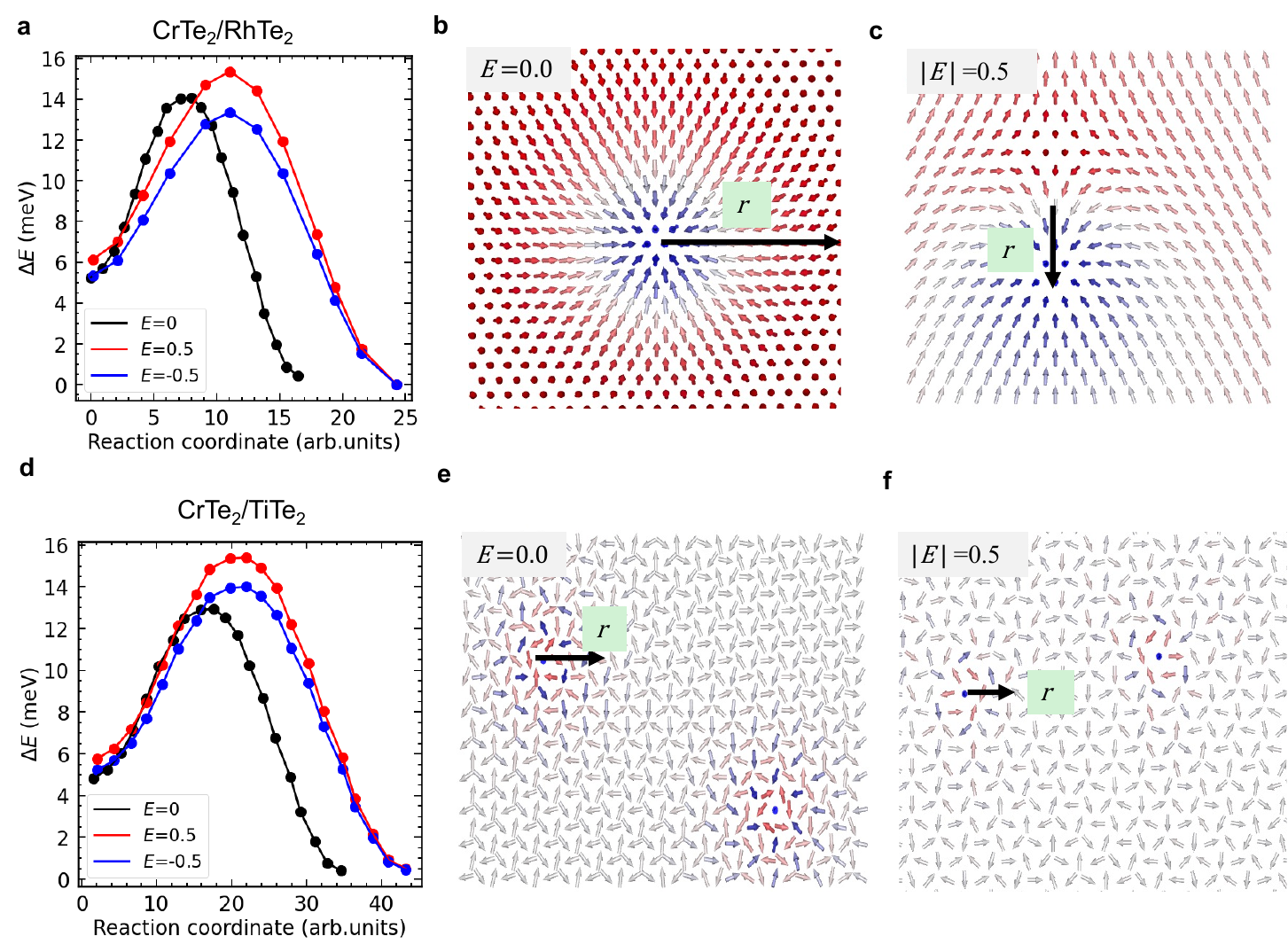}
    \caption{\label{figure_4}
    \textbf{Topological magnetic textures in heterobilayers.}
    \textbf{a} Energy path for the collapse of isolated FM skyrmions ($E = 0$) or meron pairs ($|E| = 0.5$). 
    \textbf{b} Isolated skyrmion selected to calculate energy barriers and explore its stability.
    \textbf{c} Isolated meron pair selected to calculate energy barriers and explore its stability.
    \textbf{d} Energy path for the collapse of isolated N\'eel AFM meron pairs.
    \textbf{e-f} Isolated N\'eel AFM meron selected to calculate energy barriers and explore its stability with and without electric field.
    \textbf{a-c} show results for CrTe$_{2}$/RhTe$_{2}$ and \textbf{d-f} for CrTe$_{2}$/TiTe$_{2}$, respectively.}
\end{figure}

As a final aspect, we explore whether the applied electric field can be considered for manipulation of isolated skyrmions and meron pairs in their respective magnetic backgrounds.
To do so, we perform additional spin dynamics simulations for a series of geodesic nudged elastic band (GNEB) simulations~\cite{genb}, which give access to the characteristic size of these magnetic textures as well as the energy barrier that prevents their straightforward unwinding into the respective magnetic background.
We first address CrTe$_2$/RhTe$_2$, with the computed energy barriers reported in Fig.~\ref{figure_4}a and the respective magnetic objects are illustrated in Fig.~\ref{figure_4} (b-c).
Although the electric field drastically modifies the ground state from a skyrmionic domain structure to a simpler in-plane FM, hence changing the nature of the topological defects from skyrmions to meron pairs (or in-plane skyrmions), the energy needed to create these objects from the ground state (origin of the reaction coordinate) and the respective energy barriers are quite similar.
The energy barriers show a substantial dependence on the electric field polarity ($\pm$), varying around $8.6\pm\SI{0.6}{\milli\electronvolt}$, which is a combined effect of the changes to the MAE and to the other magnetic interactions.
The radius of the meron pairs varies around $2.4\pm\SI{0.1}{\nano\meter}$ with the polarity, and they are smaller than the zero-field skyrmion which has a radius of about $\SI{3.7}{nm}$.
We next turn to CrTe$_2$/RhTe$_2$, which retains its spiralling triangular N\'eel AFM ground state when the electric field is applied.
Here the topological defects are N\'eel AFM meron pairs, with the obtained energy barriers shown in Fig.~\ref{figure_4}d and the respective magnetic structures depicted in Fig.~\ref{figure_4}e-f.
The barrier heights are $9.3\pm\SI{0.5}{\milli\electronvolt}$, which are significantly higher than the zero field value of \SI{8.1}{\milli\electronvolt}.
This is accompanied by a field-induced miniaturization of the constituent merons, with their sizes shrinking to $1.9\pm\SI{0.1}{nm}$ in comparison to the zero-field value of $\SI{2.7}{nm}$.
We attribute this shrinking to the combination of weakened Heisenberg exchange and enhanced DMI by the electric field, which enables larger angles between neighboring spin moments and so a full rotation over a smaller distance.

\section*{Discussion}

Our computational study on the electric field control of electronic and magnetic properties in CrTe$_2$/RhTe$_2$ and CrTe$_2$/TiTe$_2$ heterobilayers provides several insights into the intertwined nature of electronic, structural, and magnetic responses. 
These heterostructures are bonded by weak van der Waals interactions, against which the electric field can compete, leading to a strong dependence of the interlayer distances on the magnitude and polarity of the field.
This modulation of the interlayer distance impacts the electronic properties, specifically the charge distribution, leading to pronounced variations in local densities of states and the emergence of a screened electric field in the van der Waals gap.
It then follows that the magnetic properties are also strongly modified, evidencing a robust magneto-electric coupling. 
The Heisenberg exchange interactions, DMI values, and the magnetic ground states respond to the electric field in very distinct ways for the two heterobilayer systems, highlighting the uniqueness of each system's response.
The most striking findings are the electric-field-driven transformation of the FM skyrmion into a FM meronic state for the CrTe$_2$/RhTe$_2$ heterobilayer, while CrTe$_2$/TiTe$_2$ retains its spiralling N\'eel AFM state but with the respective AFM merons being significantly miniaturized by the electric field.

\begin{figure}[H]
    \centering
    \includegraphics[width=\textwidth]{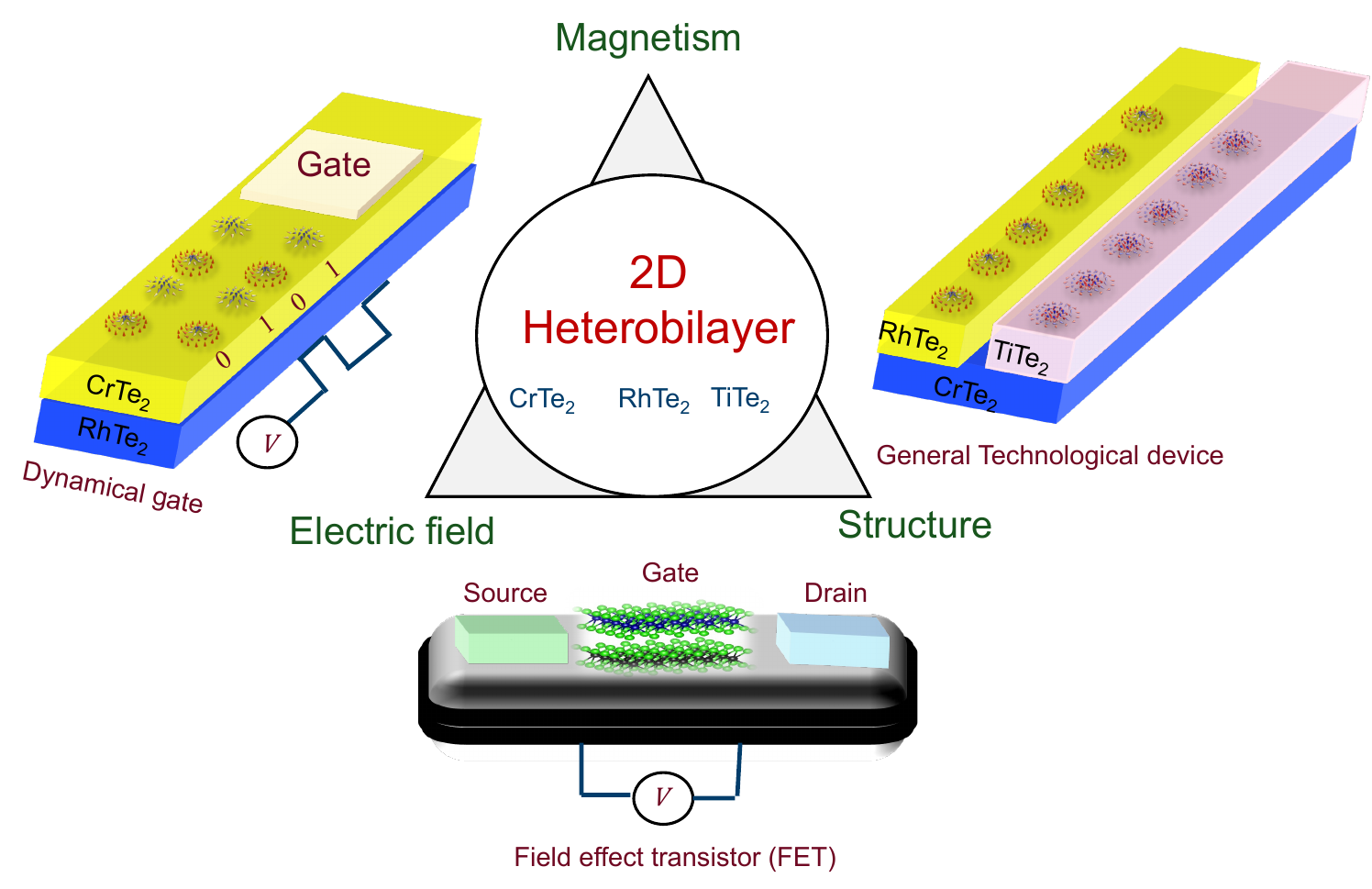}
    \caption{\label{figure_con}
    \textbf{Technological devices concepts combining 2D CrTe$_2$ with (Rh, Ti)Te$_2$}. Dynamical gate: A representation of a dynamical gate system with indications for CrTe$_2$, RhTe$_2$, and the gate voltage ($V$), which switches skyrmions into meron pairs.
    Dual racetrack: An example of a layered technological device showcasing the arrangement of CrTe$_2$, RhTe$_2$, and TiTe$_2$.
    Current injected into this device can simultaneously transport both types of magnetic objects.
    Field effect transistor (FET): A schematic representation of a FET with a focus on the gate region, indicating the source, gate, and drain alongside the voltage ($V$) applied, which controls the interlayer distance of the heterobilayer inside the gate.
  }
\end{figure}

Based on our findings, Fig.~\ref{figure_con} shows  some potential applications and implications for these observations. 
First, the ability to control the transition from FM skyrmion to FM meronic states or modulate AFM meronic configurations using electric fields can lead to new memory or logic devices, where the magnetic state (skyrmion/meron) can be considered a bit (0 or 1) and be switched electrically. 
This could also pave the way for electrically controlled computational operations, an essential requirement for scalable computing based on topological objects.
Second, the electric-field-driven modulation of the electronic properties of the heterobilayers could lead to tunable electronic devices, where the device behavior can be dynamically changed based on the applied electric field such as tunable transistors, diodes.
Finally, combining the functionalities of both FM skyrmions and AFM merons in one device could pave the way for multifunctional spintronic devices based on van der Waals heterobilayers, where data storage, transmission, logic operations, and signal processing could be integrated in a compact and efficient manner. 

\begin{methods}
\label{sec:methods}

\subsection{First-principles calculations} 

Atomic position relaxations with an electric field of CrTe$_{2}$/(Rh, Ti)Te$_{2}$ heterobilayer were assessed using density functional theory (DFT) as implemented in the Quantum Espresso (QE) computational package \cite{qe} with projector augmented plane wave (PAW) pseudopotentials \cite{ps}.
In our calculations, the generalized gradient approximation (GGA) of Perdew-Burke-Ernzerhof (PBE)\cite{PhysRevLett.77.3865} was used as the exchange and correlation functional.
The plane-wave energy cut-off is \SI{80}{\rydberg}, and the convergence criterion for the total energy is set to \SI{0.01}{\micro\rydberg}.
The self-consistent calculations were performed with a k-mesh of $24 \times 24 \times 1$ points and the Brillouin zone summations used a Gaussian smearing of \SI{0.01}{\rydberg}.
We included a vacuum region of \SI{30}{\angstrom} in the direction normal to the plane of the heterobilayer to minimize the interaction between the periodic images. 
The atomic positions were optimized by ensuring that the residual forces on the relaxed atomic positions were smaller than \SI{1}{\milli\rydberg\per\bohr}.
To study the influence of an electric field on the CrTe$_{2}$ / (Rh,Ti)Te$_{2}$ heterobilayer, a homogeneous external electric field with values changing from $\SI{0.1}{V/\angstrom}$ to $\SI{1.0}{V/\angstrom}$ was applied perpendicular to the plane of the heterobilayer.

Once the geometries of the various collinear magnetic states were established, we explored in detail magnetic properties and interactions with the all-electron full-potential relativistic Korringa-Kohn-Rostoker Green function (KKR-GF) method as implemented in the JuKKR computational package \cite{Papanikolaou2002,Bauer2014}.  
The angular momentum expansion of the Green function was truncated at $\ell_\mathrm{max} = 3$ with a k-mesh of $48 \times 48 \times 1$ points.
The energy integrations were performed including a Fermi-Dirac smearing of \SI{502.78}{\kelvin}, and the local spin-density approximation was employed~\cite{Vosko1980}.

\subsection{Magnetic interactions and atomistic spin dynamics}
\label{subsec:mi}

The magnetic interactions obtained from the first-principles calculations on the basis of the inifinitesimal rotation method \cite{inf-rot} are used to parameterize the following classical extended Heisenberg Hamiltonian with unit spins, $|\mathbf{S}| = 1$, which includes the Heisenberg exchange coupling ($J$), the DMI ($D$), the magnetic anisotropy energy ($K$), and the Zeeman term ($B$), with a finer k-mesh of $200\times 200 \times 1$:
\begin{equation}\label{eq:spin_model}
E = -\sum_i \VEC{B}\cdot \VEC{S}_i + \sum_i K_i (S_i^z)^2
- \sum_{i,j} J_{ij} \VEC{S}_i \cdot \VEC{S}_j - \sum_{i,j} \VEC{D}_{ij}\cdot(\VEC{S}_i \times \VEC{S}_j).
\end{equation}
Here $i$ and $j$ label different magnetic sites within a unit cell.
The Fourier transform of the magnetic interactions was also evaluated: $J(\VEC{q}) = \sum_j J_{0j}e^{-i\VEC{q}\cdot\VEC{R}_{0j}}$, 
where $\VEC{R}_{0j}$ is a vector connecting atoms 0 and $j$. 

Furthermore, atomistic spin dynamic simulations using the Landau-Lifshitz-Gilbert (LLG) equation as implemented in the Spirit code \cite{Mueller2019a} are performed in order to explore potential complex magnetic states while the geodesic nudged elastic band (GNEB) is utilized for investigating if these magnetic states are metastable \cite{genb, genb-1, genb-3}. 
We used the simulated annealing method: we started from a random spin state at 1000 K which we let equilibrate, then cool the system in steps by reducing
the temperature to half of its previous value and equilibrating again, until we reach
below 10 K. The value of the Gilbert damping rate in our simulation was set to 0.1.
We considered the 2D hexagonal lattice defined by the Cr atoms, with a cell size of $100 \times 100$ with periodic boundary conditions. 

\end{methods}

\noindent \textbf{Data availability}
The data that support the findings of this study are available from the corresponding author upon reasonable request.

\noindent \textbf{Code availability}
The codes employed for the simulations described within this work are open-source and can be obtained from the respective websites and/or repositories.
Quantum Espresso can be found at \cite{QEurl}, and the J\"ulich-developed codes JuKKR and Spirit can be found at \cite{JuKKRurl} and \cite{Spiriturl}, respectively.

\begin{addendum}

\item This work was supported by the Federal Ministry of Education and Research of Germany in the framework of the Palestinian-German Science Bridge (BMBF grant number 01DH16027).
We acknowledge funding provided by the Priority Programmes SPP 2244 "2D Materials Physics of van der Waals heterobilayer"  (project LO 1659/7-1) and SPP 2137 “Skyrmionics” (Projects LO 1659/8-1) of the Deutsche Forschungsgemeinschaft (DFG).
We acknowledge the computing time granted by the JARA-HPC Vergabegremium and VSR commission on the supercomputer JURECA at Forschungszentrum Jülich~\cite{jureca}. 

\item[Author contributions]
N.A. performed all the calculations, carried out the initial analysis and wrote the initial draft of the paper.
S.L. conceived, secured funding for and supervised the project.
All authors discussed the obtained results and their implications, and contributed to writing and revising the manuscript.

\item[Competing interests] The authors declare no competing interests.

\end{addendum}

\section*{References}
\bibliographystyle{naturemag}
\bibliography{Ref.bib}

\end{document}


\title{Electrical engineering of topological magnetism in two-dimensional heterobilayers}

\author[1,2,3, *]{Nihad Abuawwad}

\author[4]{ Manuel dos Santos Dias}

\author[3]{Hazem Abusara}

\author[1,2,*]{Samir Lounis}

\affil[1]{Peter Gr\"unberg Institut and Institute for Advanced Simulation, Forschungszentrum J\"ulich \& JARA, 52425 J\"ulich, Germany}
\affil[2]{Faculty of Physics, University of Duisburg-Essen and CENIDE, 47053 Duisburg, Germany}
\affil[3]{Department of Physics, Birzeit University, PO Box 14, Birzeit, Palestine}
\affil[*]{n.abuawwad@fz-juelich.de; s.lounis@fz-juelich.de}
\affil[4]{ Scientific Computing Department, STFC Daresbury Laboratory, Warrington WA4 4AD, United Kingdom}

\maketitle

\begin{figure}[H]
\centering
   \includegraphics[width=\textwidth]{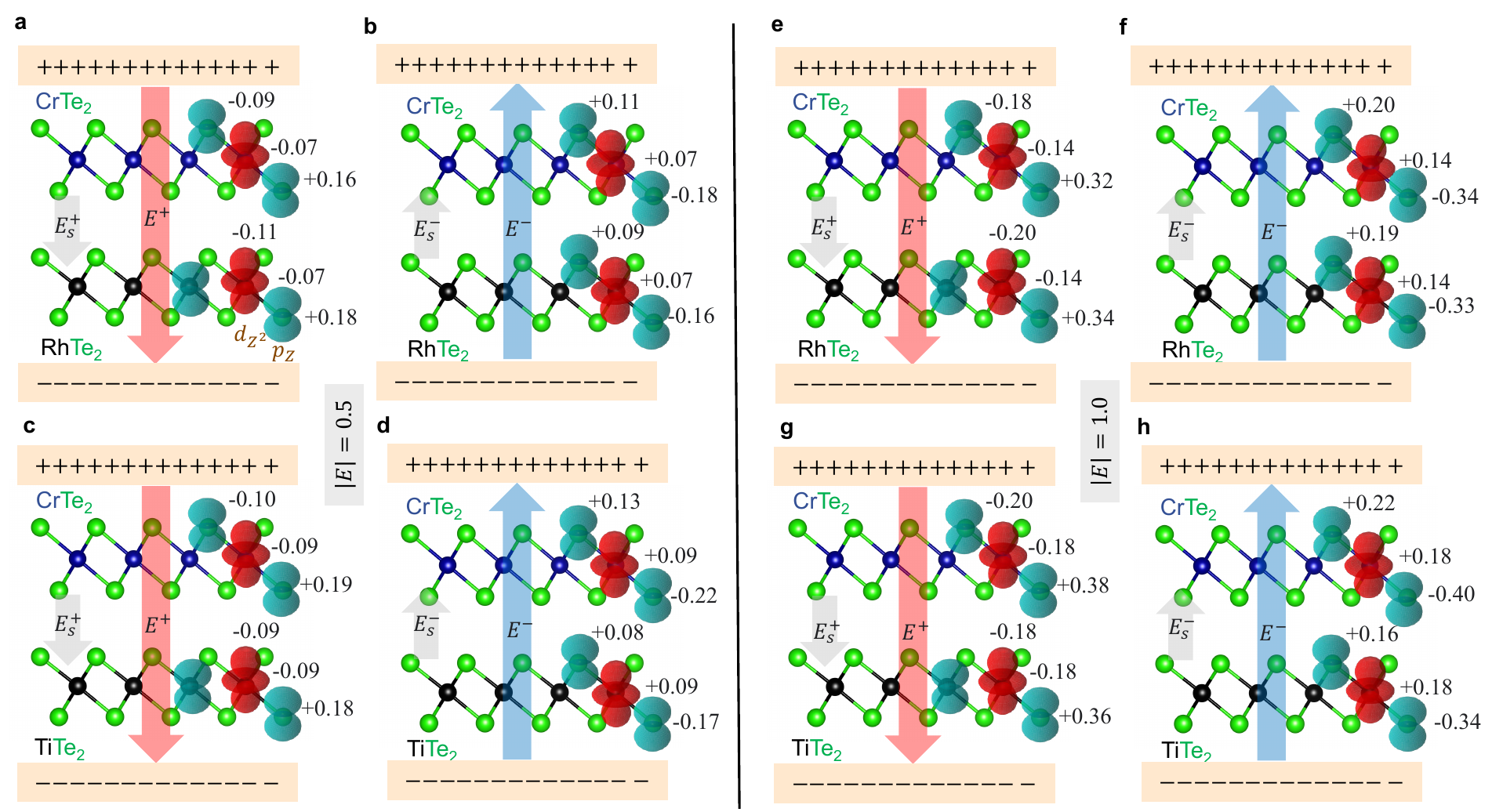}
    \caption{\textbf{Charge accumulation with respect to the perpendicular (both positive and negative) electric field.}
    $|E|$ = $\SI{0.50}{V/\angstrom}$: (a-b) CrTe$_2$/RhTe$_2$ heterobilayer. (c-d) CrTe$_2$/TiTe$_2$ heterobilayer.
    $|E|$ = $\SI{1.0}{V/\angstrom}$: (e-f) CrTe$_2$/RhTe$_2$ heterobilayer and (g-h) CrTe$_2$/TiTe$_2$ heterobilayer.
    $E_{s}^{+}$ ($E_{s}^{-}$) are the screened electric fields due to the positive (negative) external fields.}
    \label{fig-0}
\end{figure}

\begin{figure}
\centering
   \includegraphics[width=\textwidth]{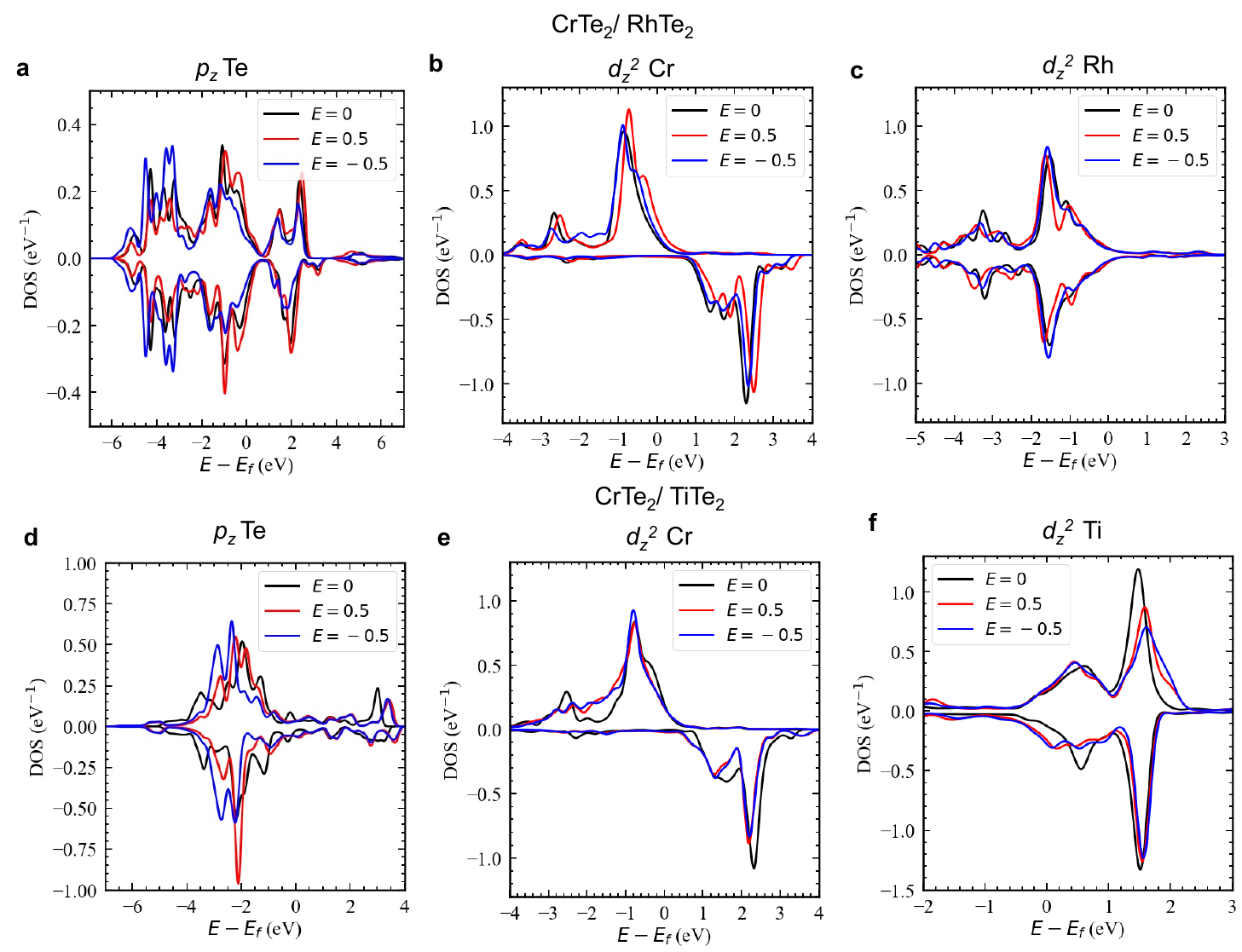}
    \caption{\textbf{Selected orbital contributions to the local density of states}.
    (a-c) CrTe$_2$/RhTe$_2$ heterobilayer.
    (d-f) CrTe$_2$/TiTe$_2$ heterobilayer.
    We plot the atom-projected contributions with $p_z$-orbital symmetry for Te, and with $d_{z^2}$-orbital symmetry for the transition metal atoms.}
    \label{fig-1}
\end{figure}

\begin{figure}
\centering
   \includegraphics[width=\textwidth]{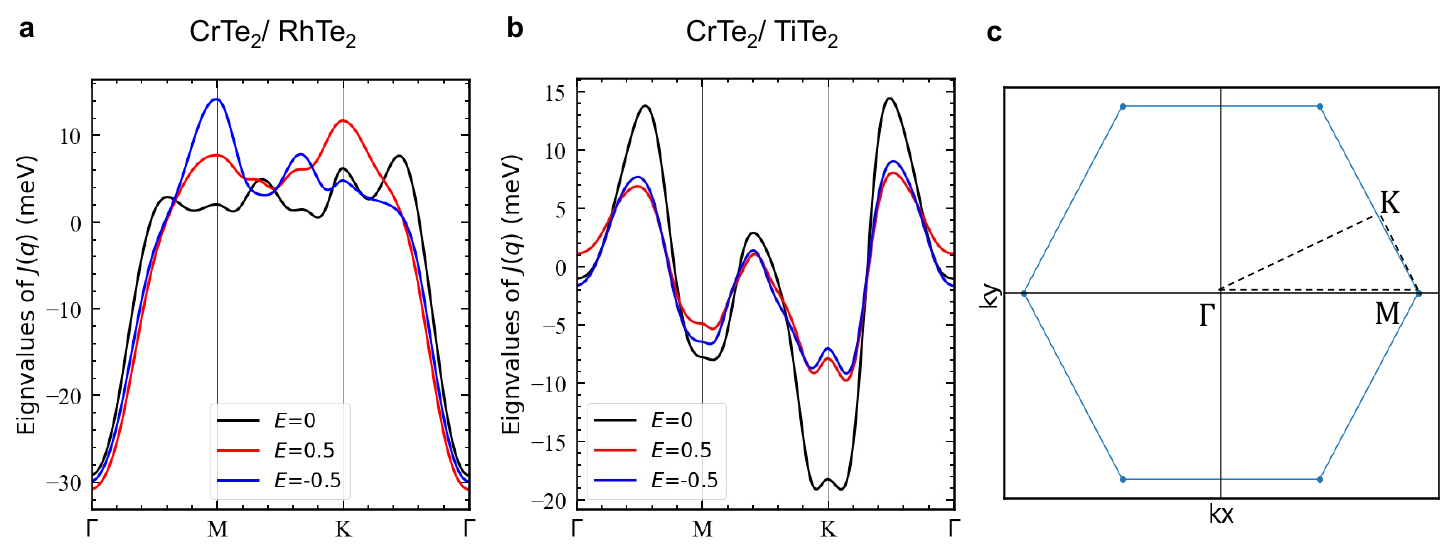}
    \caption{\textbf{Energetics of magnetic states for the heterobilayers based on the computed exchange interactions}. (a,b) Eigenvalues of the Fourier-transformed exchange interactions as a function of q at different values of the electric field in CrTe$_{2}$/RhTe$_{2}$ and CrTe$_{2}$/TiTe$_{2}$ heterobilayers, respectively. (c) The hexagonal first Brillouin zone.  }
    \label{fig-3}
\end{figure}